# BROADBAND RECEIVING SYSTEMS FOR 4.5 – 8.8 GHZ RADIO ASTRONOMICAL OBSERVATIONS AT IRBENE RADIO TELESCOPES RT-32 AND RT-16

**Vl. Bezrukovs, M. Bleiders, A. Orbidans, D. Bezrukovs**

All: Engineering Research Institute "Ventspils International Radio Astronomy Center" of Ventspils University College, Inzenieru str., 101, Ventspils, LV-3601, Latvia

**Abstract.** Since 2011 Ventspils International Radio Astronomy Centre has been involved in the large scale infrastructure project which allowed significant speeding-up of the upgrading activities related to radio telescopes RT-32 and RT-16 as to its fitting with appropriate VLBI receiving and recording equipment. Radio telescopes were instrumented with new state-of-art broadband cryogenic receivers for frequency range of 4.5 – 8.8 GHz developed and installed by company "Tecnologias de Telecomunicaciones e Informacion".

In this paper architecture of receiving system as well as significance and working principles of key subsystems are described. The receiver is formed by a cooled RF subsystem and a room temperature IF subsystem. The RF and IF subsystems are designed to process two C/X band signals (LCP & RCP) in parallel. Normally, during observations, the measured vacuum level in the receivers dewar is from $10^{-6}$ to $10^{-8}$ mbar and the temperature inside dewar is at level of 14 K at second stage, 20 K at polarizer and 46 K at the first stage.

Since October 2015 radio telescope RT-32 with new receiver system took part in several successful international VLBI sessions. During preparation for VLBI observations preliminary aperture efficiency, system temperature and beam pattern measurements were carried out to evaluate RT-32 performance after the station's renovation that besides the receiver also included repairing of the main reflector. Performance parameters were derived with the help of switching noise diode and "on-off" observations of calibration sources with known flux density at various elevations. First results measured at 4836 MHz are summarized in this manuscript.

## 1. Introduction

Ventspils International Radio Astronomy Centre (VIRAC) was established in 1994 with the aim to develop the research activities in radio astronomy and astrophysics. The instrumental base for the center comprised two fully steerable parabolic antennas, RT-16 and RT-32 (i.e. with the mirror diameter of 16 m and 32 m). The intensive reconstruction and instrumental refurbishment carried out in 2014 – 2016 made it possible to use radio telescope RT-32 for the international scale fundamental and applied research in the field of radio astronomy. The most important aspect of this work is participation of RT-32 in the VLBI (Very Long Baseline Interferometry) international experiments [1].

Since 2011 VIRAC has been involved in IKSA-CENTER Infrastructure project "Establishing National Significance Research Center for Information, Communications and Signal Processing Technologies", which allowed speeding-up of the upgrading activities related to the radio telescopes RT-32 and RT-16 as to its fitting with appropriate VLBI receiving and recording equipment. Thanks to this project, radio telescopes were instrumented with new state-of-art broadband cryogenic receivers for frequency range of 4.5 – 8.8 GHz. A set of recording equipment has already been assembled, which allows recording the signal into two channels with a bandwidth up to 1 GHz in each channel [1, 2].

One of the main scientific objectives for the VIRAC Radio astronomical observatory is VLBI (very long baseline interferometry) observations in centimeter wavelengths in collaboration with the global VLBI networks, such as European VLBI network (EVN), Low Frequency VLBI network (LFVN), and others. The new receiving and recording systems provides a high stability of the time frame, which is prerequisite for the VLBI observations. Since October 2015 radio telescope took part in several successful international VLBI sessions, and currently the priority scientific objectives for VIRAC are:
- observations in the international VLBI-networks, including the European VLBI network (EVN);
- investigation into the motion parameters of objects in the near-Earth space (space debris, satellites, asteroids) and planets;
- studying the Earth's ionosphere;
- studies of the solar radio emission.

## 2. Broadband receiving system and its subsystems

Main receiving systems for both VIRAC radio telescopes RT-32 and RT-16 are cryogenic broad band receivers for frequency range of 4.5 – 8.8 GHz developed and installed by company "Tecnologias de Telecomunicaciones e Informacion" (TTI[1]). Both receivers for RT-16 and RT-32 have identical schematics, only wideband feed-horns are different and adapted for geometry of each antenna.

This instrument is a cryogenically cooled VLBI receiver designed especially for VIRAC for astrophysical purposes. It covers the receiving frequency range from 4.5 to 8.8 GHz. The receiver is formed by a cooled RF subsystem and a room temperature IF subsystem. The RF unit amplifies signals received with wideband feed-horn which is mounted at secondary focus. RF signals are downconverted by IF unit, digitized by DBBC2 and recorded by Mark5b systems [1].

The RF and IF subsystems are designed to process two C/X band signals (LCP & RCP) in parallel, Figure 1 shows block diagram of the receiver system.

---

[1] http://www.ttinorte.es/

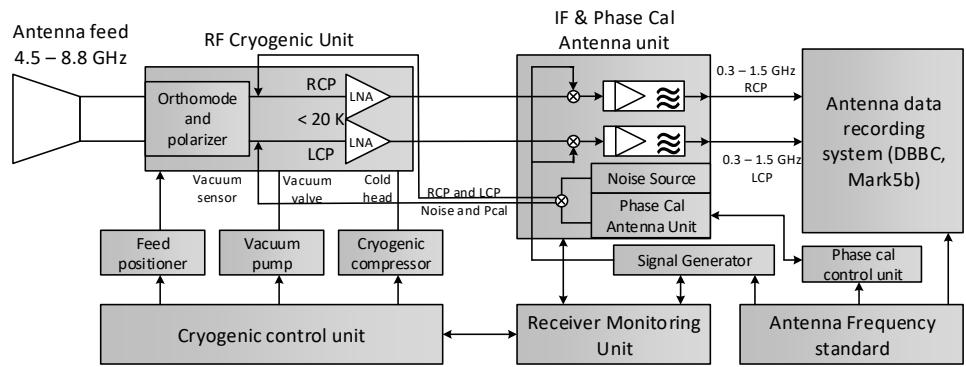

Figure 1. Block diagram of the cooled VLBI receiver system.

Physically each receiver is divided into the following subsystems:
- **Feeder System.** Receives the incoming signal from the sky. It is composed of the horn, the polarizer and orthomode transducer;
- **RF Cryogenic Unit**. Amplifies the signals coming from the feeder system. The orthomode transducer and polarizer are integrated inside the dewar of the cryostat together with the RF cryogenic unit.

Receiver control units are fitted to the 19" sub-rack:
- **Cryogenic Control Unit**. Controls the RF Cryogenic Unit and feeder positioner;
- **IF & Phase Cal Antenna Unit**. Amplifies the signals coming from the RF Cryogenic Unit and carries out the downconversion to IF band;
- **Receiver monitoring Unit**. Allows to control and monitor locally and remotely the whole receiver;
- **Phase Cal Control unit (PCCU)** together with the **Phase Cal Antenna Unit (PCAU)**. Generates the train rail of tones used to calibrate the system;
- **Standalone Signal generator.** Generates the local oscillator signal to allow the downconversion in the IF Unit.

On the Figure 2 the dewar (left) and the 19" sub-rack modules (right) installed in the RT-32 vertex rooms below the cryostat are shown.

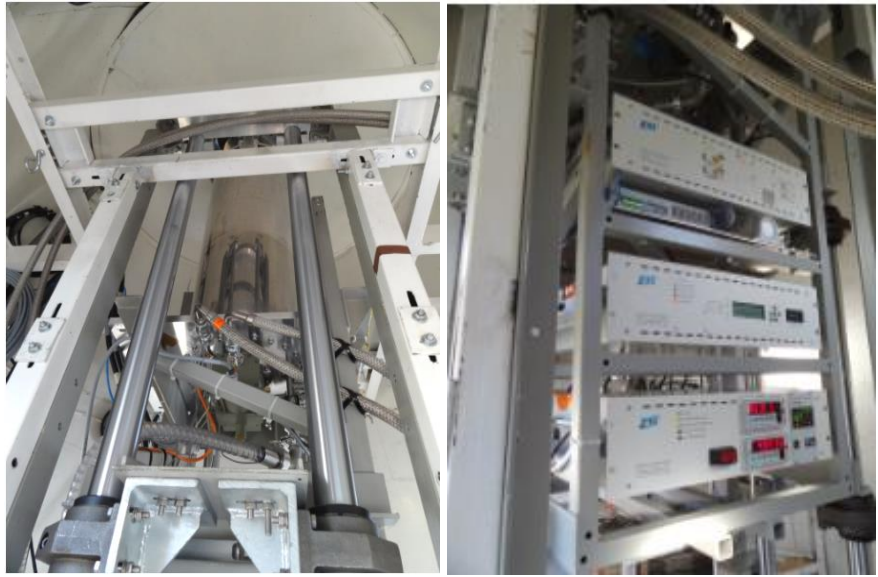

Figure 2. Receiver dewar (left); relative position of the receiver units under the cryostat (right) from top to bottom: IF & Phase Cal Antenna unit; Signal Generator; Receiver monitoring unit and Cryogenic control unit.

### 2.1. Feeder System, RF Cryogenic Unit and Cryogenic Control Unit

The receiver feeder system is composed of three elements and it is based on a passive concept:
- A primary feeder: corrugated Gaussian horn. It receives simultaneously two polarizations (LCP & RCP);
- A polarizer to convert each circular polarization into linear polarization (V & H);
- Orthomode Transducer to separate each linear polarization into a different RF path.

The feeder position can be adjusted as a function of the operating frequency with relative position shift from 0 to 600 mm.

The corrugated horn for radio telescope RT-32 is made up of five stages and a radome cover placed on top of the vertex room.

The RF unit is cooled down to cryogenic temperatures with the help of the closed-cycle Helium refrigeration system (Cryogenic system), based on the following elements:
- Air cooled single AC phase cryogenic He compressor Model 8200 Brooks-CTI;
- Vacuum subsystem based on an Adixen rotary vacuum pump model 2015-SD, automatic vacuum valve, vacuum gauge and vacuum controller from Pfeiffer. Normally, during observations, the measured vacuum level is from $10^{-6}$ to $10^{-8}$ mbar.

The temperature inside dewar is at level of 14 K (second stage), 20 K (at polarizer) and 46 K (at first stage). Receiver reaches temperature of 14 K and is ready for observations after fifteen hours of cooling. Next figures show the temperature inside the cryostat during the cooling process. After fifteen hours the temperature inside the Dewar is stabilized reaching 13 K on the cold stage, 20 K on top of the polarizer and 46 K on the intermediate stage. This measurement has been taken initially placing one of the temperature sensors on top of the polarizer and moving it towards the intermediate stage.

The cryogenic control unit implements three functions:

- Monitoring temperature inside cryostat. Two temperature sensors model DT-670 are installed inside the cryostat (Sensor A & Sensor B). Sensor A is connected to upper temperature meter in the cryogenic control unit and monitors temperature in the second stage close to the cryogenic LNAs. Sensor B is connected to lower temperature meter (Model 211S from Lakeshore) in the cryogenic control unit and monitors temperature in the first stage close to radiation screen. Both temperatures are read continuously by the Receiver Monitoring Unit;
- Monitoring of vacuum inside the cryostat. One vacuum sensor model PBR 260 from Pfeiffer is connected to the dewar. Sensor (vacuum gauge) measurement range is from $5 \times 10^{-10}$ to 1000 mbar. A vacuum meter (controller) from Pfeiffer is integrated in the Cryogenic Control Unit;
- Monitoring of heating current. Receivers include a heating subsystem of the dewar to reduce time to recover room temperature. Eight heating resistors are integrated in the dewar. Three heating resistors are connected to the first stage. Five heating resistors are connected to the second stage. Each resistor has an independent temperature relay limiting heating temperature to 30ºC. Activated from the Receiver Monitoring Unit via a local switch in the cryogenic control unit.

## 2.2. IF, Phase Cal and Noise Source subsystems

### 2.2.1. IF subsystem

After amplification with low noise cryogenic stage, RCP and LCP RF signals are downconverted using two channel IF subsystem which is integrated in IF&Phase Cal antenna unit. Block diagram of unit is shown in figure 3.

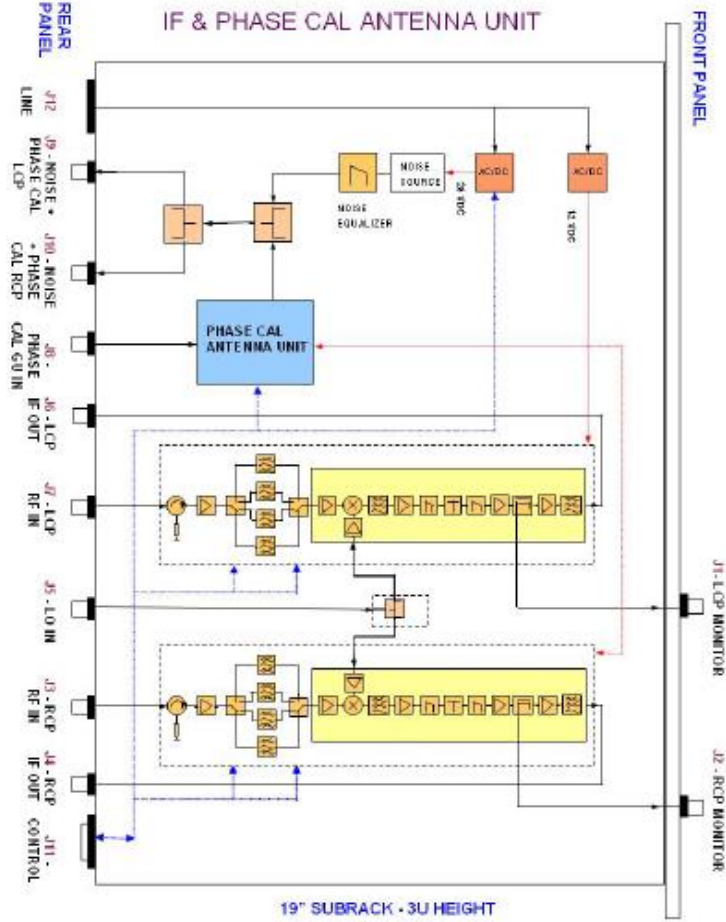

Figure 3. Detailed block diagram of IF&Phase Cal antenna unit [3].

To ensure matched load for cryogenic amplifiers, wideband isolator is used at the input of each IF cascade. Before downconversion, signal is further amplified and then filtered by four switchable image rejection bandpass filters, which are selected automatically according to user desirable frequency sub-band. These filters set instantaneously available observing bandwidth to approximately 1 GHz and provides image band rejection of more than 55 dB at all frequencies. Filtered sub-bands are then downconverted to IF frequencies in range from ≈0.3 to 1.4 GHz using wideband MMIC mixer from Hititte in conjunction with Hydrogen Maser reference locked Rhode&Schwarz SGS-B1 frequency synthesizer which has typical phase noise of -110 dBc/Hz at 10 KHz offset. Downconverted signal is low-pass filtered, further amplified. Signal spectrum also is equalized using gain slope equalizer which is tuned to compensate for attenuation slope of approximately -4.4 dB/GHz resulting at ≈65 m IF cables which carries signal from focal rooms to laboratories at ground floor of RT-32 and RT-16. As a result the gain flatness is better than +/- 1.7 dB within full 1 GHz bandpass (see figure 4). Total receiver gain of ≈94 dB is provided with 1 dB compression point power of +19 dBm at IF unit output.

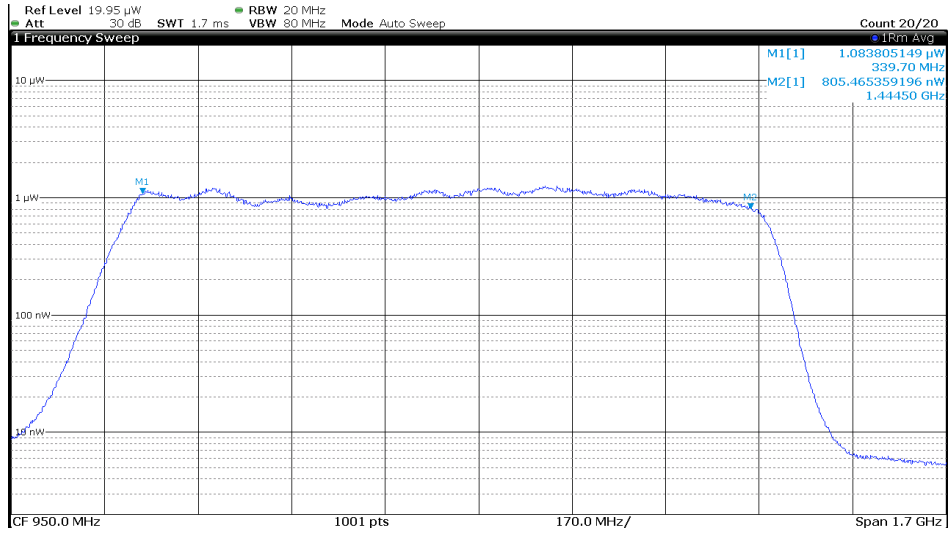

Figure 4. Example of IF signal spectrum, measured at laboratory of RT-32. LCP channel is shown.

### 2.2.2. Noise source and Phase Cal subsystems

IF&Phase Cal antenna unit also incorporates amplitude and phase calibration subsystems required for VLBI observations. Amplitude calibration is done with help of injection of noise signal with known power at the inputs of receiver. Calibration process and calculations as it is required by EVN are described in more detail in [1]. Common solid state noise source with ENR of ≈23 dB is used for calibration for both polarization channels. ENR at receiver input can be calculated by subtracting all the losses (including attenuator, cable and coupling losses) from noise source output ENR value. Expected calibration signal noise temperature then can be estimated using (1):

$$T_{cal} = 290 * 10^{\frac{ENR_{in}}{10}}, \quad (1)$$

where:
$ENR_{in}$ - excess noise ratio at receiver input, dB;
$T_{cal}$ - injected calibration signal noise temperature, K.

Value of $T_{cal}$ can be adjusted by inserting appropriate attenuator at the output of noise source. Usually a value comparable to system temperature increment caused by observable source signal is needed to be able to use same power detector gain range and reduce effective system temperature degradation in case of 80 Hz continuous calibration practiced in EVN – in case of RT-32, $T_{cal}$ should be 1 to 2 K for source flux densities on order of ~1 Jy.

One should be aware that because of large relative bandwidth, resulting frequency dependence of $T_{cal}$ value must be taken into account. Figure 5 shows calculated $T_{cal}$ versus frequency in case of different attenuator values:

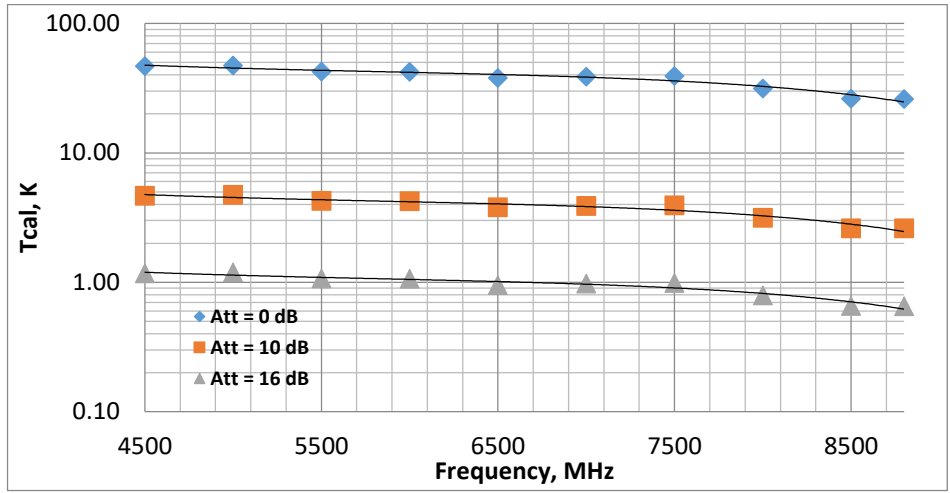

Figure 5. Frequency dependence of estimated $T_{cal}$ values in case of different noise source attenuator values.

It also should be noted that for best accuracy $T_{cal}$ must be calibrated using "hot-cold" and/or astronomical techniques and appropriate measurement bandwidths, examples of which are described in [5].

### 2.2.3. Phase Cal subsystem

VLBI observations essentially are based on time delay or phase measurement of signals received at various stations, so it is very important to compensate errors caused by station dependent phase instabilities. For such purpose receiver system at Irbene is complemented with VLBI compatible phase calibration subsch consists of temperature stabilized PCAU and PCCU. PCAU is integrated in IF&Phase Cal unit and its main purpose is generation of series of synchronized narrowband tones with 1 MHz spacing (see more details in [6]), which after summation with noise calibration signal is injected at the input of reception system (see block diagram in Figure 1). Tones can be used, for example, to compensate relative phase shifts between different baseband channels.

PCCU together with PCAU also forms maser reference signal cable delay measurement system, which is required in geodetic VLBI observations. Delay measurement is based on so called "modulated reflector" technique which eliminates errors caused by reflections – 10 MHz reference and 5 KHz tones are sent to PCAU where they are mixed together and resulting modulated reference is returned back to PCCU, which then demodulates it and derives new 10 MHz signal with phase shift proportional to two-way cable delay. Before comparing both 10 MHz signals using frequency counter, they are downconverted to 25 Hz, which enhances resolution of delay time measurement $4*10^5$ times.

### 2.3. Receiver noise temperature measurements

According to the development specifications receiver noise temperature measured at the input of the feed-horn should be below 15 K. In the table 1 noise temperature budget calculated for broad band receiver is presented.

**Table 1.** Noise temperature budget for VIRAC broad band cryogenic receivers [7].

| Component | Physical Temperature | Losses (dB) | NT (K) | Temperature | Losses (dB) | NT (K) |
|---|---|---|---|---|---|---|
| Hermetic waveguide transition | 295 K to 20K | 0.090 dB | 1.361 K | 295 K to 16K | 0.071 dB | 1.340 K |
| OMT | 20 K | 0.300 dB | 1.430 K | 16 K | 0.300 dB | 1.144 K |
| Waveguide to coaxial transition | 20 K | 0.150 dB | 0.703 K | 16 K | 0.150 dB | 0.562 K |
| Coupler | 20 K | 0.250 dB | 1.185 K | 16 K | 0.250 dB | 0.948 K |
| Coaxial Cable (cooper) TBC | 20 K | 0.194 dB | 0.913 K | 16 K | 0.194 dB | 0.731 K |
| Cryogenic Amplifier | 20 K | 32 dB | 6 K | 16 K | 32 dB | 6 K |
| Coaxial cable (cooper) | 20 K to 70 K | 0.194 dB | 2.055 K | 20 K to 70 K | 0.194 dB | 2.055 K |
| Coaxial cable | 70 K to 295 K | 0.374 dB | 16.19 K | 70 K to 295 K | 0.374 dB | 16.19 K |
| Hermetic coaxial transition | 295 K | 0.110 dB | 7.7 K | 295 K | 0.110 dB | 7.7 K |
| IF receiver NF (2.5dB) | 295 K | - | 225 K | 295K | - | 225 K |
| Total | | -30.34 dB | 11.75 K | | -30.34 dB | 10.89 K |

Figure 6 (left and right) shows results of the noise temperature measurements of actual receivers carried out at TTI premises. Y coefficient measurement technique was used with microwave absorbers at ambient and liquid nitrogen temperatures for hot and cold loads respectively. The absorbers were placed at front of the dewar RF input (non-standard waveguide transition).

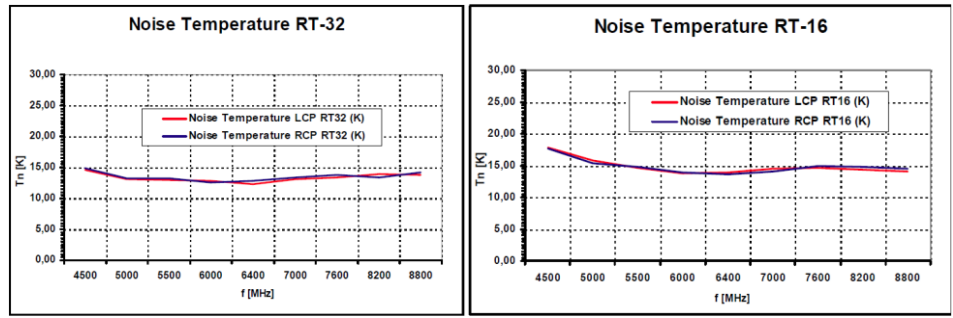

Figure 6. Results of noise temperature measurements of broadband receivers in case of radio telescopes RT-32 (left) and RT-16 (right). Measurements conducted by TTI company at TTI premises [8].

### 2.4. Remote control options for broadband receiver system

Main Control Unit (MCU) of the receiver has an optical Ethernet link to local network thus allowing remote control and monitoring of nearly all parameters of receiver. MCU accepts TCP/IP packets containing commands which should have the following structure (Figure 7):

| Start of Message | Command | Action | Data | End of Message |
|---|---|---|---|---|
| 1B | 2B | 1B | 0-1024 B | 1B |
| 0xAB | | query (?) or set (=) | | 0xBB |

Figure 7. Structure of command packet.

MCU responds to a command with acknowledge and data (if applicable) using the same structure.

There are total of 21 commands (listed in table No. 2) enabling remote setting and querying (see footnotes at the end of the table) of receiver parameters.

**Table 2.** Supported remote commands of MCU.

| | |
|---|---|
| DN | get Device Name* |
| FV | get Firmware Version* |
| OG | turn Signal Generator ON/OFF |
| AL | get Current Alarm status* |
| AH | get Alarm History (of last 30 alarms)* |
| XS | get Extended Status* |
| LN | get LNA Parameters* |
| WF | get/set Working Frequency |
| DA | turn Dry Air Control ON/OFF |
| CY | turn Cryostat Heating ON/OFF |
| RS | reset MCU |
| NS | turn Noise Source ON/OFF |
| CM | turn Compressor ON/OFF |
| VP | turn Vacuum Pump ON/OFF |
| VV | turn Vacuum Valve ON/OFF |
| SC | turn System Cooling ON/OFF |
| OG | turn Generator control ON/OFF |
| ME | Move Feeder according to WF** |
| DE | set Feeder Position to custom value / get Feeder current position |
| HE | set Feeder to Home position** |
| SE | stop Moving of Feeder** |

*-query only; **-no query possible

To query most of the receiver parameters command XS is used; it allows to get the following information (Table 3):

**Table 3.** Extended status (XS) query structure and example.

| | | |
|---|---|---|
| General structure: ¡XS=a;b;c;d;e;f;g;h;i» | | |
| Example query: ¡XS=0;+299.1;+298.7;+04.41e1;288;4F08;4500;+00.00;8800» | | |
| a | general status (e.g. ready, cooling etc.) | 0(standby) |
| b | temperature from sensor B | +299.1 K |
| c | temperature from sensor B | +298.7 K |
| d | vacuum level | +04.41e-1 |
| e | ONOFF register (indicating subsystems status) | 288 |
| f | the current alarms | 4F08 |
| g | frequency of operation | 4500 MHz |
| h | feeder position | +00.00 mm |
| i | feeder position corresponding frequency MHz | 8800 MHz |

To facilitate user interface with MCU of the receiver, graphical user interface (GUI) for Windows OS was provided by TTI. It allows easily implementing all supported remote commands of MCU as well as monitoring all the possible parameters in a single window, including local oscillator frequency which is not queryable directly.

There were two drawbacks of the GUI provided: no ability to log parameters of receiver as well as to automatically execute MCU commands (e.g. changing working frequency) according to a specified schedule. Moreover, most of RT-32 control software is being operated under UNIX platform. To solve these issues a simple command-line application `rx_main` was developed (in C programming language). The main tasks of this application are:
1) establish a TCP/IP socket connection to MCU;
2) receive any command from standard input (keyboard), form a command structure and forward it to MCU;
3) receive any response from MCU and display it on the screen;
4) implement logging of main receiver parameters.

The software performs logging each time symbol "?" is entered by user or another control software; in this case, application forwards extended status query ("`XS`") to MCU, waits for response and logs it to a file. In such a way temperature of receiver, working frequency as well as noise diode status is logged (figure 8).

```
time                 t1    t2    freq ns
2016-03-02T12:57:42 +40.5 +45.7 4836 0
2016-03-02T12:58:25 +40.6 +45.7 4836 1
```

Figure 8. Log file example of some receiver parameters.

In case of need more parameters can be logged. The application `rx_main` will be integrated in FS in the future.

There are some custom python scripts which are able to communicate with MCU via `rx_main` application. These scripts are:
1) `sched_absolute_time.py` – sends commands to MCU according to previously defined schedule file, it is used to change working frequency during observations;
2) `cal.py` – performs switching of noise diode (with command `NS`) according to the defined period; it should be noted that due to network and MCU delays command execution can be delayed either.

## 3. First results of RT-32 performance evaluation

### 3.1. RT-32 efficiency evaluation at 5 GHz

Preliminary aperture efficiency (Ap. eff.), system temperature and beam pattern measurements were carried out to evaluate RT-32 performance after the station's renovation that besides the receiver also included repairing of the main reflector. Performance parameters were carried out with the help of switching noise diode and "on-off" observations of calibration sources with known flux density at various elevations according to description in [9]. First results measured at 4836 MHz are summarized in table 4. It should be noted that elevation dependent antenna beam elevation plane offset was compensated before taking at each measurement.

**Table 4.** Summary of RT-32 performance parameters vs. various antenna elevations at 4836 MHz. Values in each row is average of multiple independent samples.

| Source | Flux, Jy | Elev., ° | $T_{sys}$ sky, K | $T_{sys}$ source, K | $T_a$, K | Ap. eff. | DPFU, K/Jy | SEFD, Jy |
|---|---|---|---|---|---|---|---|---|
| CasA | 800 | 78,00 | 30,12 | 100,74 | 70,62 | 0,30 | 0,09 | 347,3 |
| CasA | 800 | 82,00 | 27,70 | 108,69 | 80,99 | 0,35 | 0,10 | 274,4 |
| 3C123 | 16 | 24,00 | 32,82 | 34,41 | 1,59 | 0,33 | 0,10 | 353,0 |
| 3C123 | 16 | 32,00 | 38,90 | 40,76 | 1,86 | 0,40 | 0,12 | 343,1 |
| TauA | 650 | 14,00 | 41,02 | 101,41 | 60,39 | 0,32 | 0,09 | 442,6 |
| 3C454.3 | 6,45 | 47,00 | 32,70 | 34,95 | 2,25 | 0,48 | 0,14 | 239,2 |

Data shows strong $T_{sys}$ and gain (expressed in form of DPFU – degrees per flux unit) dependence on elevation which is expected. SEFD or system effective flux density integrates noise and gain performance and serves as ultimate station figure of merit and measured figures of 300 to 400 Jy is comparable with figures of other EVN stations[2]. Relatively poor aperture efficiency is expected at this stage, because required surface and secondary mirror alignments have not been attempted yet. Secondary mirror misalignment is also evident in the first results of antenna beam pattern measurements (see Figure 9). Figure 8 shows normalized DPFU dependence on elevation together with fitted second order polynomial function. It is clear that larger count of sample points is required for accurate evaluation, but it may be possible that RT-32 mirror system is optimized at elevation of approximately 45°.

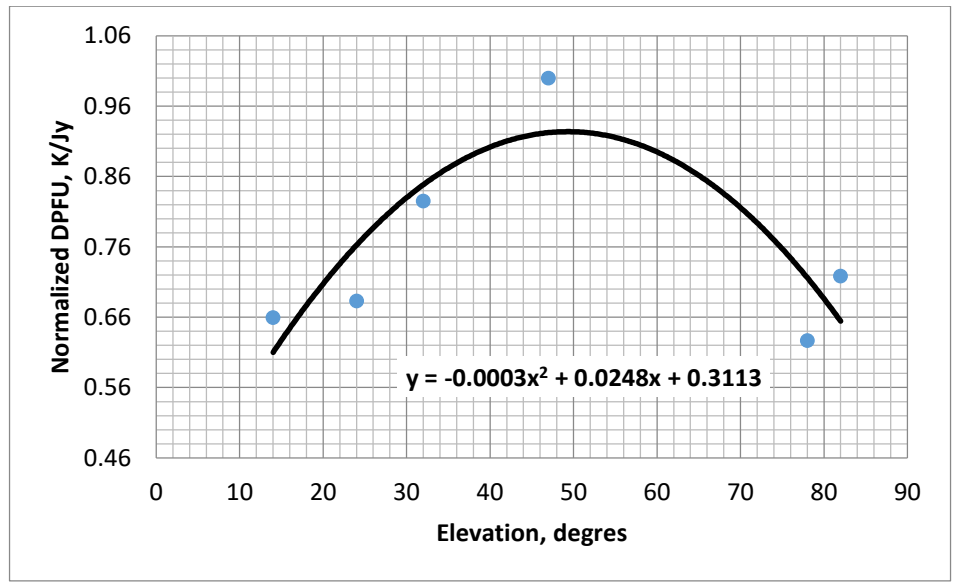

Figure 8. Degrees per flux unit (DPFU) of RT-32 versus elevation at 4836 MHz. Values are normalized to maximum DPFU of 0.14 at 47° elevation.

### 3.2. Estimation of RT-32 antenna beam pattern at 5 GHz

Preliminary test observations of bright sources had been performed to evaluate an antenna beam pattern for the corrugated C-band feed horn and an antenna coordinate frame shift from the celestial one for the new antenna drive and pointing

---

[2] http://www.evlbi.org/user_guide/EVNstatus.txt, Table II Antenna receiver performance

system. Numerous test observations of the same bright sources made before RT-32 renovation showed that the C-band antenna beam pattern had the distortion in form of two irregular side lobes [10]. This characterious distortion is caused by some radial displacement of both mirror axes and a tilt of the secondary mirror plane simultaneously. The integral power of these side lobes was about 20 – 25 % of total antenna diagram power. Therefore, the proper adjustment of the secondary mirror position should increase the effectivity of antenna up to 2-3 db.

The new antenna diagram pattern had been obtained by sequential azimuthal scanning of Cyg A source with the simultaneous elevation shift by half antenna pattern width. The antenna temperature had been recorded by total power meter and the antenna position by angle encoders had been fixed and attached to the precise time. Afterward spatial positions of antenna temperature samples was recalculated relatively the antenna geometrical axis, i.e. zeros of angle encoders.

One of the observed antenna temperature 2D distribution related to maximum is shown on Fig. 9. The approximation of the antenna temperature distribution by 2D Gaussian shows C-band HPBW about 6 – 6.5 arc. minutes taking to account the known Cyg A width at 5 cm wavelength. Obviously the center of the source is shifted from the geometrical axis of antenna and this shift has to be introduced to antenna pointing system for most effective source tracking. Two irregular antenna diagram pattern side lobes are clearly seen and the distortion remains after antenna reconstruction. So the secondary mirror adjustment is still urgent for antenna effectivity increasing.

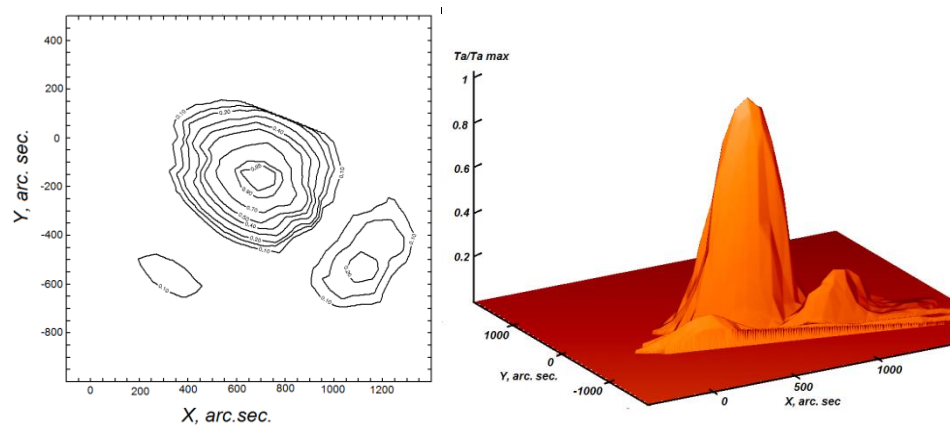

Figure 9. 2D distribution of the RT-32 C-band antenna temperature after Cyg A scanning related to the antenna geometrical axis (zeroes of angle encoders).

## 4. Conclusions

VIRAC radio telescopes RT-32 and RT-16 were instrumented with new state-of-art broadband cryogenic receivers for frequency range of 4.5 – 8.8 GHz. Since October 2015 new receiver system tested in several successful international VLBI sessions.

Physical temperature of receivers inside dewar is below 14 K at second stage, 20 K at polarizer and 46 K at first stage. Preliminary aperture efficiency, system temperature and beam pattern measurements were carried out to evaluate RT-32 performance after the station's renovation. First results measured at 4836 MHz are summarized in table 4. Data shows strong $T_{sys}$ and gain (expressed in form of DPFU – degrees per flux unit) dependence on elevation which is expected. SEFD or system effective flux density integrates noise and gain performance and serves as ultimate station figure of merit and measured figures of 300 to 400 Jy is comparable with figures of other EVN stations.

Evaluation of an antenna beam pattern at C-band shows the distortion in form of two irregular side lobes. The integral power of these side lobes was about 20 – 25 % of total antenna diagram power. The proper adjustment of the secondary mirror position should increase the effectivity of antenna up to 2 - 3 dB. Therefore, the secondary mirror adjustment is still urgent for radio telescope RT-32 effectivity increasing.

**Acknowledgements.** VIRAC radio telescopes RT-32 and RT-16 renovation and instrumentation was financed by Ventspils Town Council, Ventspils University College and IKSA-CENTER Infrastructure project "Establishing National Research center for Information, Communications and Signal Processing technologies" (Nr.2011/0044/2DP/2.1.1.3.1/11/IPIA/ VIAA/006).

This paper was supported by National research programmes' project Nr.4 "Next Generation Information and Communication Technologies (NexIT)".